\documentclass[prl,twocolumn,superscriptaddress,aps]{revtex4-1}

\usepackage{graphicx}
\usepackage{color}
\usepackage{enumerate}
\usepackage{braket}
\usepackage{amsmath,amssymb}
\usepackage{bm}
\usepackage{srcltx}

\newcommand{\pg}[1]{{#1}}

\def\g20{$g^{(2)}(0)$}

\begin{document}

\title{Strong light-matter coupling in the presence of lasing}
\author{C.~Gies}
\email{gies@itp.uni-bremen.de}
\affiliation{Institut f\"{u}r Theoretische Physik, Universit{\"a}t Bremen, 28334 Bremen, Germany}
\author{F.~Gericke}
\affiliation{Institut f\"{u}r Festk\"{o}rperphysik, Technische Universit{\"a}t Berlin, Hardenbergstra{\ss}e 36, 10623 Berlin, Germany}
\author{P. Gartner}
\affiliation{National Institute of Materials Physics, P.O.Box MG-7, Bucharest-Magurele, Romania}
\author{S.~Holzinger}
\affiliation{Institut f\"{u}r Festk\"{o}rperphysik, Technische Universit{\"a}t Berlin, Hardenbergstra{\ss}e 36, 10623 Berlin, Germany}
\author{C.~Hopfmann}
\affiliation{Institut f\"{u}r Festk\"{o}rperphysik, Technische Universit{\"a}t Berlin, Hardenbergstra{\ss}e 36, 10623 Berlin, Germany}
\author{T.~Heindel}
\affiliation{Institut f\"{u}r Festk\"{o}rperphysik, Technische Universit{\"a}t Berlin, Hardenbergstra{\ss}e 36, 10623 Berlin, Germany}
\author{J.~Wolters}
\affiliation{Institut f\"{u}r Festk\"{o}rperphysik, Technische Universit{\"a}t Berlin, Hardenbergstra{\ss}e 36, 10623 Berlin, Germany}
\author{C.~Schneider}
\affiliation{Technische Physik, Universit{\"a}t W{\"u}rzburg, Am Hubland, 97074 W{\"u}rzburg, Germany}
\author{M.~Florian}
\affiliation{Institut f\"{u}r Theoretische Physik, Universit{\"a}t Bremen, 28334 Bremen, Germany}
\author{F.~Jahnke}
\affiliation{Institut f\"{u}r Theoretische Physik, Universit{\"a}t Bremen, 28334 Bremen, Germany}
\author{S.~H\"{o}fling}
\affiliation{Technische Physik, Universit{\"a}t W{\"u}rzburg, Am Hubland, 97074 W{\"u}rzburg, Germany}
\affiliation{SUPA, School of Physics and Astronomy, University of St. Andrews, St. Andrews KY16 9SS, United Kingdom}
\author{M.~Kamp}
\affiliation{Technische Physik, Universit{\"a}t W{\"u}rzburg, Am Hubland, 97074 W{\"u}rzburg, Germany}
\author{S.~Reitzenstein}
\affiliation{Institut f\"{u}r Festk\"{o}rperphysik, Technische Universit{\"a}t Berlin, Hardenbergstra{\ss}e 36, 10623 Berlin, Germany}

\date{\today}

\begin{abstract}
The regime of strong light-matter coupling is typically associated with weak excitation. With current realizations of cavity-QED systems, strong coupling may persevere even at elevated excitation levels sufficient to cross the threshold to lasing. In the presence of stimulated emission, the vacuum-Rabi doublet in the emission spectrum is modified and the established criterion for strong coupling no longer applies. We provide a generalized criterion for strong coupling and the corresponding emission spectrum, which includes the influence of higher Jaynes-Cummings states. The applicability is demonstrated in a theory-experiment comparison of a few-emitter quantum-dot--micropillar laser as a particular realization of the driven dissipative Jaynes-Cummings model. Furthermore, we address the question if and for which parameters true single-emitter lasing can be achieved, and provide evidence for the coexistence of strong coupling and lasing in our system in the presence of background emitter contributions.
\end{abstract}


\maketitle

\section{I. Introduction}
Strong coupling (SC) and lasing are usually observed in disjunct operational regimes: SC in the quantum regime for weak excitation of a single emitter \cite{khitrova_vacuum_2006}, and lasing using strong excitation of many emitters, such as an ensemble of atoms, or a semiconductor gain material, in the weak-coupling regime \cite{ulrich_photon_2007}. The regime where both effects intermingle is widely unexplored, but has stirred interest since it was first reported in a semiconductor nanolaser \cite{nomura_laser_2010}. SC is generally identified by the occurrence of two well-separated
peaks in the emission spectrum, as defined by analytic expressions known from textbooks (e.g. Ref.~\cite{carmichael_statistical_2007}), in which the spectrum is written as modulus square of the difference between two poles,
\begin{equation}
\begin{aligned}
\label{eq:Scarmichael}
S(\omega) \sim \left|\frac{1}{\omega-\omega_1} - \frac{1}{\omega-\omega_2}\right|^2 \, .
\end{aligned}
\end{equation}
For a discrete emitter, this is the so-called vacuum Rabi doublet--- the
spectral representation of vacuum-Rabi oscillations that arises from the coherent energy exchange
between light and matter
\cite{khitrova_vacuum_2006,carmichael_statistical_2007}. In the presence of
dephasing, which originates from cavity losses, spontaneous emission, and
carrier relaxation processes following excitation, it is known that strong
coupling  persists as long as $4g > |\Gamma-\kappa|$, where $g$ is the
light-matter coupling strength, $\kappa$ the cavity loss rate, and $\Gamma$
gives the total exciton dephasing, before the Rabi doublet merges into a
single line marking the transition to weak coupling \cite{carmichael_statistical_2007}.

In this letter, we show that this well-established criterion and the emission spectrum \eqref{eq:Scarmichael} with
poles defined as in \cite{reithmaier_strong_2004,carmichael_statistical_2007} fail to describe the system close to the lasing
threshold due to the onset of stimulated emission. At stronger excitation, contributions from higher
excited states begin to matter, and their influence also affects the vacuum Rabi doublet. This effect is in addition to previous analytic considerations of emission spectra in driven dissipative systems \cite{del_valle_regimes_2011,del_valle_luminescence_2009,laussy_strong_2008,yao_nonlinear_2010}. By extending the Hilbert space beyond commonly used approximations, we provide a
correction term to the strong-coupling criterion that shifts the strong-to-weak
coupling transition to significantly lower excitation powers in agreement with a
full non-perturbative solution of the driven dissipative Jaynes-Cummings model.


Our work relates to the quantum regime of SC, which is between discrete states of the quantized light field and an electronic transition \cite{khitrova_vacuum_2006}, such as realized in atoms \cite{bernardot_vacuum_1992}, superconducting circuits \cite{wallraff_strong_2004,wang_measurement_2008}, and semiconductor quantum dots (QDs) \cite{yoshie_vacuum_2004,reithmaier_strong_2004} coupled to micro- or nanocavities. Their small mode volume facilitates single-mode lasing with
only few discrete emitters, allowing to push the concept of lasing into the domain of quantum optics. In particular, we study the ultimate limit of a single-QD laser \cite{noda_seeking_2006} that has been pursued \cite{xie_influence_2007,reitzenstein_single_2008,nomura_photonic_2009} in analogy to the single-atom laser \cite{mckeever_experimental_2003}. In these systems important questions remain, such as to the influence of non-resonant background emitters \cite{winger_explanation_2009,ates_non-resonant_2009,chauvin_controlling_2009, laucht_temporal_2010}. By combining spectroscopy and autocorrelation measurements with density-matrix calculations, in our joint experimental and theoretical work we address the following questions: Can a single
quantum emitter provide sufficient gain to achieve lasing? Which signatures can be used to uniquely identify the transition from strong to weak coupling? Can SC and lasing coexist? Answering these questions will strongly advance our knowledge at the quantum level of light-matter interaction with important interdisciplinary impact in quantum optics, nanophotonics and in the development of ultimate nanolasers.

This paper is structured as follows: In the next section (II) we derive the analytic spectrum for different approximations and define strong-coupling conditions.
In Sec. III the conditions for single-QD lasing and strong-coupling are discussed.
Sec. IV presents the experimental setup and the measurements and compares the theoretical findings to the experimental results. Sec. V concludes this work.

\section{II. SC condition and emission spectrum}
The underlying quantum-mechanical problem is defined by the model of a two-level emitter coupled to
a cavity photon mode through the Jaynes-Cummings (JC) interaction as described by the Hamiltonian
\pg{(in $\hbar=1$ units)}
\begin{equation}
H = H_0 + H_{JC} = \omega_c\,c^\dagger c +\omega_0\,b^\dagger b
   + g\,(b^\dagger v^\dagger c + b\, c^\dagger v) \, .
\label{eq:ham}
\end{equation}
The operators $b$ and $b^\dagger$ refer to cavity-mode photons and we use QD notations for the two-level emitter: $c, c^\dagger$ are the fermionic annihilation and creation operators for carriers occupying the upper (conduction band) level and $v, v^\dagger$ for the lower (valence band) one, whose energy is taken as zero.

%
To describe a driven, dissipative laser system, the Hamiltonian dynamics
is augmented by dissipative processes and pumping via various Lindblad
terms acting on the density operator $\rho$ as
${\cal L}_X [\rho] = \frac{1}{2} \Gamma_X \left\{2\,X\,\rho\, X^\dagger -
X^\dagger X \,\rho - \rho\,X^\dagger X \right\}$, with $\Gamma_X$ the corresponding rate.
The time evolution of the density operator is given by the von Neumann-Lindblad
(vNL) equation
\begin{equation}
\frac{\partial}{\partial t}\,\rho = -i\left[H, \rho \right] + \sum_X {\cal
L}_X [\rho] \, ,
\label{eq:vNL}
\end{equation}
and the incoherent processes considered in the summation above are
(i) cavity losses, defined by $X=b$ with rate $\Gamma_b$ denoted as
$\kappa$, (ii) exciton decay with $X=v^\dagger c$ and rate
$\Gamma_{v^\dagger c}=\gamma$, (iii) pumping, represented by an
up-scattering process with $X= c^\dagger v$, $\Gamma_{c^\dagger v} = P$, and
(iv) pure dephasing leading to homogeneous line broadening, defined by
$X=c^\dagger c$ with the rate $\Gamma_{c^\dagger c} = \gamma_h$.

In a rotating frame picture defined by $\widetilde H_0 = \omega_b(c^\dagger c +
b^\dagger b)$, and writing explicitly the Lindblad contribution of the
incoherent processes defined before, the equation of motion (EoM) for the
expectation value of an arbitrary operator $A$ reads
\begin{align}
\frac{\partial}{\partial t}\,\Braket{A}
= & -i \Braket{\left[ A,\,\Delta \,c^\dagger c + g\,(b^{\dagger}v^\dagger c
     + b\,c^\dagger v)\right ]} \nonumber \\
  & + \frac{\kappa}{2}\,\Braket{\left [b^\dagger ,A \right]\, b +
     b^\dagger \left[A, \,b \right]} \nonumber \\ \,
  & + \frac{\gamma}{2}\,\Braket{\left [c^\dagger v,\,A \right]\, v^\dagger c+
     c^\dagger v \left[A,\,v^\dagger c \right]} \, , \nonumber \\
  & + \frac{P}{2}\,\Braket{\left [v^\dagger c,A \right]\, c^\dagger v+
     v^\dagger c \left[A,\,c^\dagger v \right]} \, ,\nonumber \\
  & + \frac{\gamma_h}{2}\,\Braket{\left [c^\dagger c,\,A \right]\, c^\dagger c+
     c^\dagger c \left[A,\,c^\dagger c \right]} \, .
\label{eq:eom}
\end{align}
Note that the last term can be written in several equivalent ways, for instance
with all the $c$-operators replaced by $v$-operators, or as $\gamma_h/4 \cdot
\Braket{\sigma_z A\, \sigma_z-A}$, see e.g. \cite{auffeves_controlling_2010}.

An analytic expression for the emission spectrum, such as
Eq.~\eqref{eq:Scarmichael}, and the criterion for SC can only be
obtained by using approximations that limit the Hilbert space to a
low-excitation subspace.
We compare two approximations: (i) the \emph{three-state approximation} (3SA), which
reproduces the well-known expression \eqref{eq:Scarmichael} \pg{with poles
differing from Ref.~\cite{carmichael_statistical_2007} by taking into account
pure and excitation-induced dephasing} \cite{del_valle_luminescence_2009,yao_nonlinear_2010}.
This approximation considers only states not exceeding a
total excitation of $N_\mathrm{ex}=c^\dagger c+b^\dagger b= 1$.
Explicitly, these are the ground state $\Ket{v,0}$ and the
states with one excitation $\Ket{c,0}$ and $\Ket{v,1}$. (ii) the \emph{four-state approximation} (4SA) is derived by including the additional state $\ket{c,1}$ with $N_\mathrm{ex}=2$ in the derivation, providing
corrections that improve the description of systems driven close to the laser threshold.


The cavity emission spectrum is calculated using the first order
auto-correlation function of the photon operators
\begin{equation}
g_b^{(1)}(t) =  \lim_{t' \to \infty}\Braket{b^\dagger(t')\,b(t+t')} \, .
\end{equation}
The long time limit implies that the expectation values are calculated using
the steady-state (ss) density operator $\rho_{ss}$, and the correlation function
is {\it formally} the expectation value of $b(t)$ using an auxiliary "density
operator" $\rho_b$:
\begin{equation}
g_b^{(1)}(t) = \Braket{b(t)}_b =  \text{Tr}
\left \{\rho_b\, b(t) \right \} \, ,
\qquad  \rho_b = \rho_{ss}\, b^\dagger \, .
\end{equation}
The emission spectrum is then given by the expression
\begin{equation}
S_b(\omega) = 2 \,\text{Re} \int_0^\infty g_b^{(1)} (t)\, e^{i \omega t}
\text{d} t = 2\,\text{Re}\, g_b^{(1)}(\omega) \, .
\label{eq:def_spec}
\end{equation}
The EoM for the evolution of the correlation function $\left< b(t) \right>_b$
is obtained from the same Eq.~(\ref{eq:eom}), as for any operator expectation
value (Quantum Regression Theorem) \cite{gardiner_quantum_2004}, irrespective of the density operator
involved in the averages. It is by the initial conditions alone that the
solution depends on the particular density operator considered.
In the present case the initial conditions are expressed as expectation
values on the steady-state density operator
$\left< A(0)\right>_b = \text{Tr} \left\{\rho_{ss}b^\dagger A\right\}=
\left< b^\dagger A\right>_{ss}$.

\subsection{A. Low-excitation approximations}

The EoM for the quantity of interest $\Braket{b(t)}_b$ generates a hierarchy of
equations for higher operator averages \cite{gies_semiconductor_2007, kira_semiconductor_2011}. As mentioned above, the form of these
equations is independent on the density operator, therefore we drop in the
following the sub-index $b$. Also, for simplicity the time argument is left
out. One obtains successively
\begin{subequations}
\begin{align}
\frac{\partial}{\partial t}\Braket{b}
   &= -\,ig \Braket{v^\dagger c} -\frac{\kappa}{2}\Braket{b} \, ,
       \label{eq:hchy_a} \\
\frac{\partial}{\partial t}\Braket{v^\dagger c}
   &= - \left(\frac{\widetilde P}{2} +i\Delta\right)\Braket{v^\dagger c} +
      ig\Braket{b \left( c^\dagger c - v^\dagger v \right) } \, ,
        \label{eq:hchy_b} \\
\frac{\partial}{\partial t}\Braket{b\,c^\dagger c}
   &= -\,ig\Braket{b\,b\,c^\dagger v}
   + ig \Braket{b^\dagger b\,v^\dagger c} \notag\\
   & \quad -\left(\gamma
      +\frac{\kappa}{2} \right)\Braket{b\,c^\dagger c}
      + P \Braket{b\, v^\dagger v} \, ,
          \label{eq:hchy_c} \\
\frac{\partial}{\partial t}\Braket{b^\dagger b\,v^\dagger c}
   &= - \left(\frac{\widetilde P + 2 \kappa}{2} +i\Delta \right)
       \Braket{b^\dagger b\,v^\dagger c} \notag\\
       & \quad +ig\Braket{b^\dagger b\,b
       \left(c^\dagger c - v^\dagger v\right)} +ig \Braket{b\,c^\dagger c} \; ,
       \label{eq:hchy_d}
\end{align}
\label{eq:hchy}
\end{subequations}
where by $\widetilde P$ we denoted  $P +\gamma+\gamma_h$. One can eliminate
averages containing the $v^\dagger v$ operator in favor of $c^\dagger c$ using
$v^\dagger v = 1 - c^\dagger c$.

The chain of EoM is infinite, involving growing products of operators. In order
to obtain a finite, closed set of equations some approximations are needed. If
the system is not strongly pumped it is natural to limit the Hilbert space of
the problem to the low excited states. This can be done in several ways, as
seen in what follows.

\subsubsection {1. The three-state approximation (3SA)}\label{sec:3SA}
The ground state of system consisting of the emitter plus cavity mode, as
described by $H_0$ of Eq.~\eqref{eq:Scarmichael}, is the state $\Ket{v,0}$ with the
emitter in its lower state and no photon in the cavity. No excitation is
present in the system. The states with one excitation are $\Ket{c,0}$ and
$\Ket{v,1}$, and the vacuum Rabi oscillation is the energy exchange between
these two. Limiting the Hilbert space to these three states, i.e. to the states
with no more than one excitation is the approximation considered here (3SA).

In this case it is easy to see that $\Braket{b\,c^\dagger c}$ can be discarded,
as it requires more than one excitation. Thus in Eq.~\eqref{eq:hchy_b} one has
$\Braket{b\,\left( c^\dagger c - v^\dagger v \right)} = -\Braket{b}$, and
Eqs.~(\ref{eq:hchy_a}),(\ref{eq:hchy_b}) become a closed set of two equations
for $\Braket{b}$, and $\Braket{v^\dagger c}$. The two-dimensional  evolution
problem has the form
\begin{equation}
\frac{\partial}{\partial t}\Ket{\Psi}= M \Ket{\Psi}\, ,
   \qquad \Ket{\Psi} = \left|\begin{array}{c}
                        \Braket{b} \\
                        -ig\Braket{v^\dagger c}
                      \end{array}
                                 \right> \, ,
\label{eq:2D_Psi}
\end{equation}
and the Fourier transform of the time evolution, required by
Eq.~\eqref{eq:def_spec}, amounts to a matrix inversion problem
\begin{multline}
 \int_0^\infty \Ket{\Psi(t)} e^{i\omega t} \text{d} t\\
= \int_0^\infty e^{\,(i\omega+M)t}\Ket{\Psi(0)} \text{d} t =
-(i\omega+M)^{-1}\Ket{\Psi(0)} \, .
\label{eq:ft_psi}
\end{multline}
In our case the matrix to be inverted is
\begin{equation}
\left( \begin{array}{cc}
         i\omega-\kappa/2 & 1  \\ \\
            -\,g^2   & \quad i\omega' -\widetilde P/2
           \end{array} \right)
 =  \left( \begin{array}{cc}
         D_1(\omega) & 1  \\ \\
            -\,g^2   & D_2(\omega)
           \end{array} \right)
\label{eq:2D_M}
\end{equation}
where $\omega'=\omega-\Delta$ and $\widetilde P = P+\gamma+\gamma_h$.

The inverse is given by
\begin{equation}
(i\omega +M) ^{-1} = \frac{1}{\text{det}(i\omega+M)}
       \left( \begin{array}{cc}
        D_2(\omega)  & -\,1  \\ \\
        g^2   & \quad D_1(\omega)
           \end{array} \right)
           \, ,
\label{eq:2D_res}
\end{equation}
where the determinant of $i\omega+M$ is
\begin{equation}
D(\omega)= D_1(\omega)\, D_2(\omega)+g^2 =
\left(i\omega-\frac{\kappa}{2}\right)\cdot
\left(i\omega'- \frac{\widetilde P}{2} \right) + g^2 \, .
\label{eq:D_om}
\end{equation}
The matrix of Eq.~\eqref{eq:2D_res} should be applied to the vector of initial
conditions and the result projected on the first component, corresponding to
$\Braket{b}$
\begin{equation}
g^{(1)}(\omega)=  - \left<1\, , 0 \right| (i\omega+M)^{-1}
    \left | \begin{array}{c}
            \Braket{b^\dagger b} \\
             -ig \Braket{b^\dagger v^\dagger c}
             \end{array}
             \right>_{ss} \, .
\label{eq:g1_om}
\end{equation}
The last step is the calculation of the steady-state expectation values
defining the initial conditions. To this end one has to examine the chain of
EoM associated with the photon number
\begin{subequations}
\begin{align}
\frac{\partial}{\partial t}\Braket{b^\dagger b}
 &= \phantom{-} 2\, \text{Re}\{-ig\Braket{b^\dagger v^\dagger c}\}
   - \kappa\, \Braket{b^\dagger b}
   \label{eq:ss_a} \\
 \frac{\partial}{\partial t}\Braket{c^\dagger c}
 &= - 2\,\text{Re}\{-ig\Braket{b^\dagger v^\dagger c}\}
    -\gamma \,\Braket{c^\dagger c} + P\, \Braket{v^\dagger v}
   \label{eq:ss_b} \\
\frac{\partial}{\partial t}\Braket{b^\dagger v^\dagger c}
 &= - (\Gamma/2 + i \Delta) \Braket{b^\dagger v^\dagger c} \notag \\ &\quad
  +ig \Braket{b^\dagger b(c^\dagger c - v^\dagger v)}
  +ig \Braket{c^\dagger c} \, ,
    \label{eq:ss_c}
\end{align}
\label{eq:ss}
\end{subequations}
where $\Gamma= \widetilde P+\kappa = P + \gamma +\gamma_h + \kappa$ sums up all
the dephasing processes. Here again the chain is broken by limiting the Hilbert
space to the subspace with no more than one excitation. Indeed, in this case
$\Braket{b^\dagger b\,c^\dagger c}=0$ and therefore $\Braket{b^\dagger
b\,v^\dagger v}= \Braket{b^\dagger b}$ and one is left with only three unknowns:
the photon number $N= \Braket{b^\dagger b}$, the exciton population $n=
\Braket{c^\dagger c}$ and the photon assisted polarization $\varphi = -ig
\Braket {b^\dagger v^\dagger c}$. Moreover, in the steady state the time
derivatives are zero and one is left with a system of three algebraic equations
for these unknowns. One obtains for the steady-state values
\begin{equation}
\varphi = \frac{2\,g^2}{\Gamma+2\, i\Delta}\,(n-N) \,,
\label{eq:phi}
\end{equation}
with $R$ having the familiar expression for the spontaneous emission rate.
Eventually one finds
\begin{align}\label{eq:phieq2supp}
N &= \frac{R\, P}{R (P+\gamma+\kappa)+\kappa (P+\gamma)} \nonumber \\
\varphi &=\frac{2\,g^2}{\Gamma+2\, i\Delta}\,\frac{\kappa}{R} \,N
  = \frac{\kappa}{2} \left(1-2\,i\frac{\Delta}{\Gamma} \right)  N \, .
\end{align}
Collecting all these results one obtains
\begin{equation}
g^{(1)}(\omega)= \frac{-D_2(\omega)N + \varphi}{D(\omega)} = \frac{\Gamma/2
-i\omega'-i\Delta\,\kappa/\Gamma}{D(\omega)}\, N \, .
\label{eq:2D_g1_om}
\end{equation}
The factor $N$ is frequency independent and thus is not influencing the shape
of the spectrum. Its presence is related to our choice of the normalization of the correlation function $g^{(1)}$. Therefore one can simplify both Eq.~\eqref{eq:2D_g1_om} and the definition of $\varphi$ in Eq.~\eqref{eq:phieq2supp} by setting $N=1$ without influencing the spectral lineshape.

The important feature here is the position of the poles of $g^{(1)}(\omega)$,
i.e. the zeroes of $D(\omega)$. These are easily found analytically as
the roots of a second degree polynomial. The peaks of $S(\omega)$, measured from
the cavity frequency $\omega_b$, are located at the real parts of these roots.

As an example we consider the resonant case $\Delta = 0$, in which the roots
$\omega_{1,2}$ of $D(\omega)$ are given by
\begin{equation}
\omega_{1,2} = -i\frac{\Gamma}{4} \pm \sqrt{g^2-\left(\frac{\widetilde P
-\kappa}{4} \right)^2} = i\frac{\Gamma}{4} \pm g' \, .
\end{equation}
Obviously, the existence of two distinct peaks at $\omega=\pm g'$ is
conditioned by $g'$ being real, or
\begin{equation}
4\,g > \left|\widetilde P -\kappa \right|\, .
\end{equation}
According to Eq.~(\ref{eq:def_spec}) the spectrum is given, up to a
normalization factor, by
\begin{equation}
S(\omega) =
\frac{(\Gamma/2-i\omega)D^*(\omega)+(\Gamma/2+i\omega)D(\omega)}{
D(\omega)D^*(\omega)} \, .
\label{eq:spec_3lea}
\end{equation}
With $D(\omega)= -\omega^2 - i\omega\,\Gamma/2 + g^2+\widetilde P\, \kappa/4$
it is immediate that the numerator of Eq.~(\ref{eq:spec_3lea}) does not depend
on $\omega$ and therefore the shape of the spectrum is given by
\begin{equation}
S(\omega) \sim \left|\frac{1}{D(\omega)} \right|^2 \sim
\left|\frac{1}{\omega-\omega_1} - \frac{1}{\omega-\omega_2}\right|^2 \, .
\end{equation}
This expression for the spectral shape is similar to the one
derived in \cite{carmichael_subnatural_1989, carmichael_statistical_2007},
which is also obtained using only the three lowest-excited states. Our result is
slightly more general, as it includes the presence of pumping.

\subsubsection {2. The four-state approximation (4SA)}\label{Sec:4SA}

Instead of limiting the Hilbert space to states with {\it up to one excitation}
one can consider the subspace with {\it up to one photon}, which means
taking into account a fourth state, namely $\Ket{c,1}$. This improves the
approximation without including a higher rung of the JC ladder, so that it
still deals only with the vacuum Rabi oscillations.

Now the expectation value $\Braket{bc^\dagger c}$ is not discarded from the
picture and additional EoM have to be considered. It is the terms containing
$b\,b$ in Eqs.~(\ref{eq:hchy_c}, \ref{eq:hchy_d}) which vanish, since
they require two photons to be annihilated. As a consequence the whole set of
equations Eqs.~(\ref{eq:hchy}) is now a closed system for four unknowns. We
choose them as $\Braket{b}$, $-ig\Braket{v^\dagger c}$, $\Braket{b\,c^\dagger
c}$, and $-ig\Braket{b^\dagger b\, v^\dagger c}$ and denote the four-dimensional
vector having these components by $\Ket{\Psi}$. Its time evolution is generated
by a four-dimensional matrix M. As before, we need the inverse of a matrix,
which now has the form
\begin{equation}
i\omega +M   = \left( \begin{array}{cccc}
   D_1(\omega) & 1  & 0  & 0 \\
 - g^2 & D_2(\omega) & 2\,g^2 &  0\\
   P   & 0   & D'_1(\omega) & -1\\
   0   & 0   & g^2& D'_2(\omega)
            \end{array}
    \right) \, ,
    \label{eq:4D_M}
\end{equation}
with $D'_1(\omega)=i\omega-(P+\gamma+\kappa/2)$,
$D'_2(\omega)= i\omega'-(\widetilde P/2\,+\kappa)$ and $D_1(\omega),
D_2(\omega)$ as previously defined. Considering the matrix as split into 2x2
blocks, the upper-left one is the same as discussed above in 3SA. The
lower-right block is quite similar, with the determinant given by
$D'(\omega)=D'_1(\omega)D'_2(\omega) +g^2$. The simple, block-diagonal picture
is perturbed by the presence of the off-diagonal blocks. The latter are sparse,
so that the total determinant can be easily calculated
\begin{equation}
\text{det}(i\omega+M) = D(\omega)\cdot D'(\omega) +2\, g^2 P D'_2(\omega) \, .
\end{equation}
The expression of $g^{(1)}(\omega)$ is a four-dimensional analog of
Eq.~(\ref{eq:g1_om})
\begin{equation}
g^{(1)}(\omega)=  - \left<1,\, 0\,, 0\,, 0 \right| (i\omega+M)^{-1}
\Ket{\Psi(0)} \, .
\end{equation}
This time the components of $\Ket{\Psi(0)}$ are the steady-state values of
$N=\Braket{b^\dagger b}$, $\varphi=-ig \Braket{b^\dagger v^\dagger c}$,
$K=\Braket{b^\dagger b\, c^\dagger c}$ and $\lambda =-ig \Braket{b^\dagger
b^\dagger\,b\, v^\dagger c}$, calculated in the 4SA. It is immediate that the
last component $\lambda$ is zero in this approximation, so that the calculation
of $g^{(1)}(\omega)$ involves only three matrix elements of the cofactor of
$i\omega+M$. One obtains
\begin{equation}
g^{(1)}(\omega) = \frac{\left[-D_2(\omega)N+\varphi \right] \,D'(\omega)- 2\,
g^2D'_2(\omega)K}{D(\omega)\cdot D'(\omega) +2\, g^2 P D'_2(\omega)} \, .
\label{eq:4D_g1_om}
\end{equation}
It is obvious that the last terms in both the numerator and denominator of
Eq.~(\ref{eq:4D_g1_om}) make the difference between 4SA and 3SA. Without them
one recovers the result of Eq.~(\ref{eq:2D_g1_om}).

Having included the fourth state $\Ket{c,1}$, the system of Eqs.~(\ref{eq:ss})
is not closed anymore and has to be supplemented. Indeed, now
$K=\Braket{b^\dagger b\, c^\dagger c}$ cannot be discarded and its EoM has to
be added
\begin{multline}
\frac{\partial}{\partial t}\Braket{b^\dagger b\, c^\dagger c}= 2\,\text{Re}
\left\{ig\,\Braket{b^\dagger b^\dagger b\, v^\dagger c} \right\} \\
-(\gamma+\kappa)\Braket{b^\dagger b\, c^\dagger c} +P \Braket{b^\dagger b\,
v^\dagger v} \, .
\label{eq:ss_d}
\end{multline}
Here the first term is negligible, since it contains two photonic creation
operators and the system becomes closed. Its solution in the steady state
now reads:
\begin{align}
N & =\frac{R\,P}{R\,(P+\gamma+\kappa)+\kappa\,(P
+\gamma)-2\,R\,P(P+\gamma)/(P+\gamma+\kappa))} \nonumber \\
\varphi & = \frac{\kappa}{2}\,\left(1-2\,i\frac{\Delta}{\Gamma}\right)\, N
\nonumber \\
K & = \frac{P}{P+\gamma+\kappa}\, N \label{eq:supp_varphi}
\end{align}
The photon number $N$ is modified with respect to its 3SA value by the last
term in the denominator. It should be noted that the above expression for $N$
coincides with the lowest truncation of its continued fraction expansion
[$r_1 = 0$ in Eq.~(12) of \cite{gartner_two-level_2011}. As in the 3SA case, $N$ plays the role of a normalization constant, and can be taken equal to 1 both in Eq.~\eqref{eq:4D_g1_om} and in the expressions for $\varphi$ and $K$ in Eq.~\eqref{eq:supp_varphi}.

The positions of the spectral peaks are given by the roots of the
denominator of Eq.~(\ref{eq:4D_g1_om}), which is now a four-degree
polynomial. In the limit $P \to 0$ two of the four zeroes are the roots of
$D(\omega)$ as in 3SA, while the other two are new and correspond to the
zeroes of $D'(\omega)$. Since they evolve continuously with increasing $P$, one
can trace back which of them started as roots of $D(\omega)$ and which stem from
the new roots. We call the former the "main" roots since it turns out that the
spectrum is essentially determined by them. The other, "secondary" roots give
rise to small corrections. Their contribution is not even systematically
positive, so there is no {\it bona fide} spectrum associated with them.
\begin{figure}[]
 \centering
  \includegraphics[width=\columnwidth]{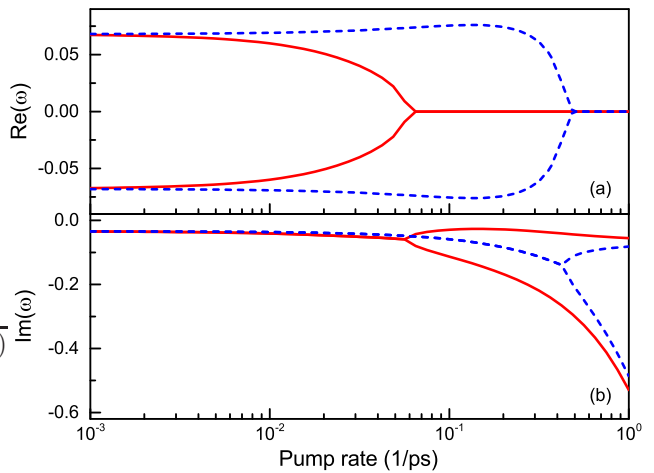}
  \caption{(Color online) (a) The real and (b) imaginary parts of the
two main roots (solid, red line) obtained by 4SA. For comparison the 3SA roots are also shown (dashed, blue line). It is seen that the merging of the
peaks occurs earlier in 4SA than in 3SA. The parameters used are $\Delta=0$,
$\kappa= 0.136$, $g =0.076$, $\gamma= 0.1$, $\gamma_h=0$, all in $ps^{-1}$
units.}
  \label{fig:root_cm}
\end{figure}
It should be noted that in the resonant case the denominator of
Eq.~(\ref{eq:4D_g1_om}) is a polynomial with real coefficients in the argument
$z=i \omega$, and therefore the roots $z$ are either real or pairwise complex
conjugated. In the latter case, in the $\omega$ plane the roots have the real
parts equal and of opposite sign, and the imaginary parts coincide. This is
seen in Fig.\ref{fig:root_cm} for low pump values, and corresponds to the
strong coupling regime. Later the real parts merge in zero and the imaginary
parts start taking different values (weak coupling).

It is seen that new dephasing terms, contained in $D'_1(\omega)$ and
$D'_2(\omega)$, are bigger than those in $D_1(\omega)$ and $D_2(\omega)$,
because a larger product of operators in the expectation values comes with
stronger dephasing. As a consequence the pumping interval of strong coupling is
expected to be overestimated by the 3SA. This is confirmed by
Fig.\ref{fig:root_cm}.

Focusing now on the main poles, it would be desirable to obtain them at least
approximately as the roots of a second degree polynomial. This would not
only simplify the search for their positions but would also allow a more direct
comparison with the 3SA result and point out the correction terms. We describe
below a scheme for reaching this aim, in the case of resonance.

\begin{figure}[]
 \centering
  \includegraphics[width=\columnwidth]{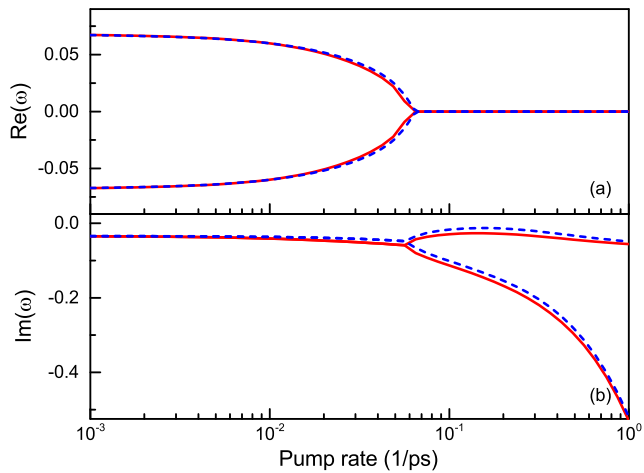}
  \caption{(Color online) (a) The real and (b) imaginary parts of the
two main roots (solid, red line) obtained by 4SA, in comparison with
the roots of Eq.~\eqref{eq:simfit}. The
parameters are the same as in Fig.\ref{fig:root_cm} (dashed, blue line).}
  \label{fig:root}
\end{figure}

To this end we rearrange the equation
\begin{equation}
D(\omega)\cdot D'(\omega) +2\, g^2 P D'_2(\omega) = 0 \, ,
\end{equation}
in a way that separates it into the 3SA denominator $D(\omega)$ plus a
"correction"
\begin{equation}
D(\omega) +\frac{2\, g^2 P D'_2(\omega)}{D'(\omega)} = 0 \, .
\end{equation}
This suggests a self-consistent scheme, in which the argument $\omega$
in the correction term is a constant updated at each iteration. As the
starting point one may choose for this constant the value of the average $\bar
\omega = -i\Gamma/4$ of the 3SA roots. The resulting second degree polynomial
is
\begin{multline}
\chi(\omega) = D(\omega) +\frac{2\, g^2 P D'_2(\bar\omega)}{D'(\bar\omega)} \\
= \left [ \left(i \omega - \frac{\kappa}{2} \right)\cdot\left(i \omega -
\frac{\widetilde P}{2} \right) +g^2 \right] \\
 - 2 g^2 P\, \frac{\widetilde P/2
+\kappa - \Gamma/4}{(P + \gamma+\kappa/2 -\Gamma/4)(\widetilde P/2
+\kappa -\Gamma/4)+g^2} \, ,
\label{eq:simfit}
\end{multline}
and its roots already provide a good approximation for the main roots of 4SA,
as seen in Fig.\ref{fig:root}. Therefore there is no need for additional
iterations. Of course, the accuracy of the approximation might depend on the
parameters and a careful examination of the various cases should be performed.
The correction introduced in Eq.~(\ref{eq:simfit}) goes in the direction of
replacing $g^2$ by a smaller quantity, and thus it reduces the domain of strong
coupling.

%


\begin{figure}
\centering
\includegraphics[width=1\columnwidth]{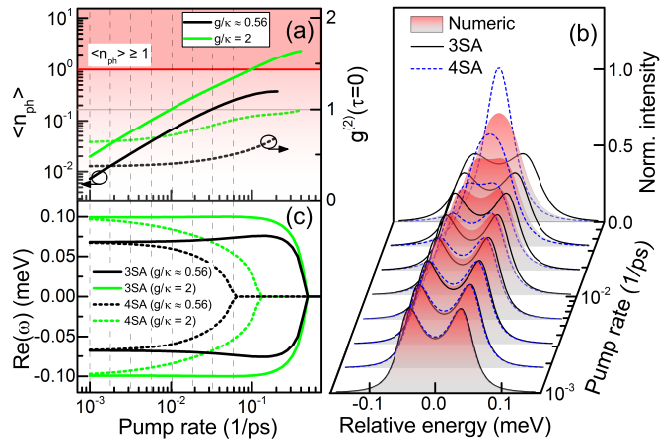}
\caption{\textbf{(a)} Input-output curve and \textbf{(b)} calculated cavity emission spectra for a single-QD microlaser with the parameters of the experiment (black: $\Delta=0$,
$\kappa= 0.136$, $g =0.076$, $\gamma= 0$, $\gamma_h=0$ in ps$^{-1}$) for the pump rates indicated by the vertical lines in (a) and (c).
The approximate analytical 3-state approximation (3SA) (solid line) and 4SA (dashed line) results discussed in the text are compared to the numerical spectra (shaded area) obtained by solving Eq.~\eqref{eq:vNL}. All shown spectra are normalized to unity area.
\textbf{(c)} Real parts of the main roots of $g_b^{(1)}(\omega)$ indicating the transition from strong to weak coupling.
SC persists for all in (b) shown spectra according to the 4SA condition. In \textbf{(a)} additional results for a single-QD-laser that overcomes the threshold are shown in green ($\kappa= 0.05$, $g =0.1$, $\gamma= 0$, $\gamma_h=0$ in ps$^{-1}$).
}
\label{fig:SQDlaser}
\end{figure}

\section{III. Condition for single-QD lasing and SC}
It has been widely discussed that a determination of the laser transition in high-$\beta$ lasers is difficult from the input-output curve alone \cite{chow_emission_2014}, and the photon autocorrelation function $g^{(2)}(0)$ is generally used to identify thermal, coherent, or single-photon emission in terms of $g^{(2)}(0)=2,1$ and $<0.5$, respectively. In Fig.~\ref{fig:SQDlaser}(a) we show input-output curve and $g^{(2)}(0)$ (black curves) for a single-QD microcavity system with $g/\kappa \approx 0.5$, which corresponds to our experiment and is a value also realized in other studies on QD-microcavity systems \cite{majumdar_cavity_2012,lermer_bloch-wave_2012,nomura_laser_2010}. As can be seen, the emission saturates before lasing is reached due to the limited gain that the single emitter can provide. To attain $g^{(2)}(0)\approx 1$ and $\langle n_\mathrm{ph}\rangle=1$ requires at least $g/\kappa > 2$ for the single emitter (green curves). While such a high values may be realized via further technological improvements in terms of ultra-high cavity Q-factors and significantly larger light-matter coupling constants \cite{ota_spontaneous_2011}, this agrees with previous predictions that with dielectric cavity designs, a single-QD contributes significantly to lasing but additional background gain is required to reach and overcome the laser threshold \cite{gies_single_2011, reitzenstein_single_2008}.

The cavity emission spectra corresponding to the black curves in (a) are shown in Fig.~\ref{fig:SQDlaser}(b) and reveal a  transition from a doublet to a single-peak structure. The merging of the peaks in the full numerical solution of Eq.~\eqref{eq:vNL} (shaded) is well reproduced by the 4SA Eq.~\ref{eq:4D_g1_om} (dashed lines), while the 3SA (solid lines) fails to correctly predict this behavior within the investigated excitation range. From Eq.~\eqref{eq:4D_g1_om} we can directly determine the transition from strong to weak coupling. The real part of the poles of $g_b^{(1)}(\omega)$ is shown in Fig.~\ref{fig:SQDlaser}(c). The 4SA (dashed curve) predicts the transition to take place at a pump rate that is nearly one order of magnitude lower compared to the conventional 3SA-criterion $4g> |\widetilde P - \kappa|$, with $\widetilde P$ the total exciton dephasing in our case (solid curve).

\begin{figure}[t!]
\centering
\includegraphics[width=1\columnwidth]{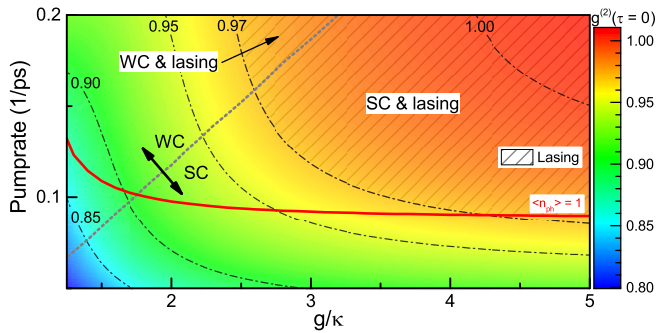}
\caption{Parameter-space diagram relating the conditions for SC (dotted line) and lasing (red: $\langle n_\mathrm{ph}\rangle=1$, $g^{(2)}(0)$ as colormap) to the dimensionless light-matter coupling and excitation strength.}
\label{fig:phasespace}
\end{figure}

More insight is obtained from a parameter-space map that shows the three criteria discussed above in terms of the key system parameter $g/\kappa$ that determines the existence and perseverance of strong light-matter coupling, and excitation strength $P$ that enters into $\Gamma$ as excitation-induced dephasing. 
The line that separates the strong and weak coupling regime as defined by the merging of the two main spectral poles of Eq.~\eqref{eq:def_spec}
is represented in Fig.~\ref{fig:phasespace} by a dotted line. The red line marking the pump rate at which $\langle n \rangle=1$ and $g^{(2)}$-values as colormap are superimposed. As a criterion for lasing, we here use a 3\% margin around $g^{(2)}=1$ as indicated by the contour line. Four regimes can be distinguished, in which either SC, lasing, neither, or both is realized. Lasing with only a single-QD is possible for $g/\kappa \gtrsim 2.5$. At these large coupling strengths, lasing takes generally place in the presence of SC, and lasing in the weak-coupling regime is only realized if the excitation power is increased further than the threshold value.  For $g/\kappa \lesssim 2.5$ SC of a single emitter and lasing can coexist if the missing gain is provided by additional background emitters, as we now discuss in the context of experimental results.


%

\section{IV Theory-experiment comparison}

\subsection{1. Sample fabrication and experimental setup}
The QD-micropillar laser is based on a planar microcavity structure grown by molecular beam epitaxy (inset of Fig.~\ref{fig:Sample}(a)). The central layer is composed by a single layer In$_{0.4}$Ga$_{0.6}$As quantum dots located in the center of one-$\lambda$ thick GaAs cavity layer. We have chosen laterally extended QDs with an Indium content of about $40\%$ and an area density of $10^{10}$~cm$^{-2}$ to foster pronounced cavity-QED effects in the single-QD regime.
On top (bottom) of the GaAs cavity $26$ ($30$) pairs of AlAs/GaAs layers act as highly reflective distributed Bragg-reflecting mirrors. The investigated micropillar with a diameter of $1.8~\mu$m and a quality-factor of $Q=15,000$ was realized by high-resolution electron-beam lithography and plasma etching \cite{reitzenstein_quantum_2010}. 

Optical studies were performed at cryogenic temperatures using a Helium flow cryostat with a standard high resolution confocal micro-photoluminescence ($\mu$PL) setup. The measured signal was collected by an objectiv with a numerical aperture of 0.4 and dispersed by a spectrometer with a resolution of $25~\mu$eV and a fiber-based Hanbury-Brown and Twiss (HBT) configuration with two different sets of single-photon counting modules with a total temporal resolution of about 500\,ps (60\,ps) and a high (low) quantum efficiency. A frequency-doubled Nd:YAG-Laser at 532\,nm in continuous wave (cw) mode was used for optical excitation.

\subsection{2. Excitation-power dependence of the QD-micropillar emission}
\begin{figure}[t!]
\centering
\hspace*{-7mm}
\centering \includegraphics[width=\columnwidth]{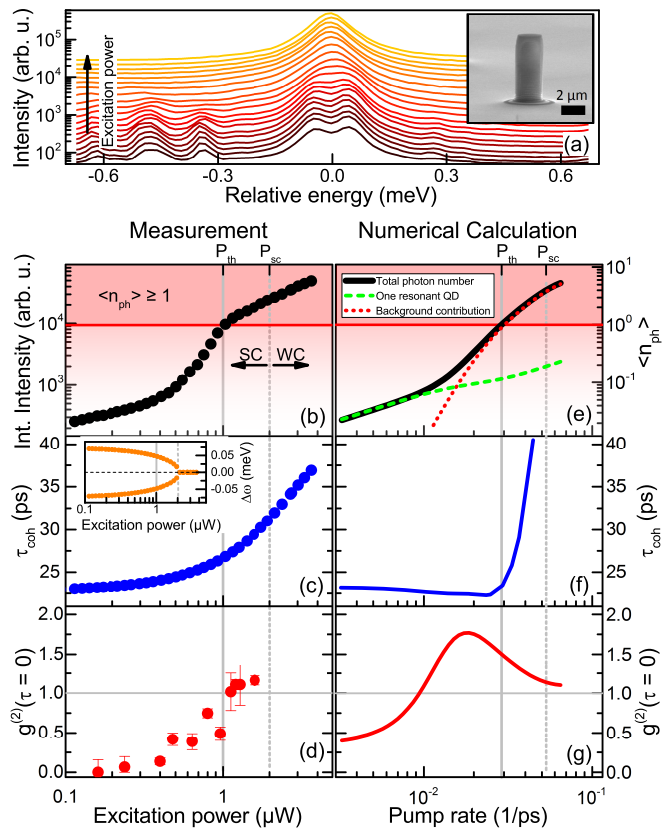}
\caption{ \textbf{(a)} Excitation power series of $\mu$PL spectra recorded at resonance. Scanning electron micrograph of a micropillar with a diameter of 2\,$\mu$m shown as inset.
\textbf{Bottom:} Laser characteristics obtained from experiment (\textbf{left}) and theory (\textbf{right}). From top to bottom, input-output curve, coherence time and second-order photon correlation function are shown, in their combination providing evidence that the microlaser crosses the transition to lasing. Experimental coherence times have been obtained by using Eq.~\eqref{eq:4D_g1_om} to model the measured spectra, followed by Fourier transform and integration $\tau_\mathrm{coh} = \int \mathrm{d}\tau |g^{(1)}(\tau)|^2$. \textbf{Inset:} Position of the roots of Eq.~\eqref{eq:simfit} calculated for the corresponding fit parameters.
}
\label{fig:Sample}
\label{fig:Setup}
\label{fig:TSerie}
\label{fig:exc-dep_spec_EXP}
\label{fig:thexp}
\end{figure}

The excitation-power dependent evolution of the emission spectra in Fig.~\ref{fig:thexp} (a) demonstrates the disappearance of the vacuum Rabi doublet into a single emission peak with increasing excitation power suggesting a transition into the weak coupling regime. In addition to the Rabi doublet, emission from three non-resonant QDs can be seen at negative detuning (at around $-0.5$\,meV). It is commonly agreed that the dephasing associated with the scattering grows with increasing carrier density \cite{uskov_line_2001} and is the origin of the line broadening that ultimately causes the transition to weak coupling \cite{carmichael_statistical_2007,munch_role_2009,favero_temperature_2007}. By fitting the experimental emission spectra using Eq.~\eqref{eq:4D_g1_om} for a fixed set of parameters, taking only into account the respective excitation power, we can directly evaluate the SC criterion \eqref{eq:simfit}. The real part of the roots of $\chi(\omega)$ are shown as inset to panel (c) of Fig.~\ref{fig:thexp} and reveal that the transition to weak coupling takes place at about $2\,\mu$W.

In the following we identify the lasing characteristics of our device by a combined experimental and theoretical analysis of the emission intensity, autocorrelation function and coherence time as function of pumping. In Fig.~\ref{fig:thexp}(b)--(d) experimental data is shown. 
From the autocorrelation measurements we identify the regime where the single QD-gain contribution clearly dominates the emission, indicated by $g^{(2)}(\tau=0)<0.5$ ($P<1\mu$W). On the other hand, at high excitation powers a continuous increase of the output intensity is observed in (b), which is a signature of background contributions instead of that of a single, saturable emitter. We explain this transition by additional emitters present in the cavity. At low excitation, their excitonic transitions are detuned from the cavity mode. At intermediate excitation levels around $1\,\mu$W, multi-exciton transitions become realized and spectrally overlap with the cavity mode \cite{laucht_temporal_2010}, providing cavity feeding that first leads to not yet fully coherent emission with photon bunching ($g^{(2)}(0) > 1$) before the threshold is crossed and emission becomes fully coherent ($g^{(2)}(0)\approx 1$). An observed increase in coherence time (c) is also indicative for the onset of lasing \cite{chow_emission_2014}. The coherence time is obtained by using Eq.~\eqref{eq:4D_g1_om} to fit the emission spectra for a single set of parameters (only $P$ is variable) followed by Fourier transformation. This parameter set is then used in the theoretical calculation.

The above discussion on the interplay of single-QD and background contributions is the foundation for our theoretical modelling of the few-emitter nanolaser. Most published work including \cite{nomura_laser_2010} use a phenomenological photon-generation term by means of inverse cavity losses to account for background effects, which has the difficulty that photons are purely thermal \cite{florian_phonon-mediated_2013,gies_single_2011}. Instead, we treat background emitters on a microscopic footing by solving Eq.~\eqref{eq:vNL} directly for a few-emitter system and input parameters taken from the experiment. Due to the complexity of the calculation, we treat the single-QD and background-dominated excitation regimes separately: In the low-excitation regime ($P<0.01/$ps) Eq.~\eqref{eq:vNL} is solved for a single emitter, whereas a single QD plus up to seven transitions of background emitters are explicitly included at higher excitation ($P>0.02/$ps), when higher multi-exciton states acting as gain centers become realized with sufficient likelihood.
The transition depends on the exact mechanism of the non-resonant coupling, for which we use a fit as it is not a focus of this work. By including all contributing emitters in Eq.~\eqref{eq:vNL}, we are able to correctly account for the properties of the gain material and, thereby, obtain realistic linewidth and $g^{(2)}$-values in the presence of background effects.

Theoretical results are shown Fig.~\ref{fig:thexp}(e)--(g). The calculated input-output curve first shows a linear increase, which arises from the exciton transition of the single QD (contribution marked in green). When the exciton of the single QD saturates, multi-exciton states of the background emitters begin to add to the photon emission into the mode (their contribution is marked in red). The kink in the input-output curve is, therefore, not related to the $\beta$ factor (for the strongly-coupled QD we assume $\beta\approx 1$), but arises from the transition from single-QD to background-dominated emission. Lasing with $g^{(2)}(0)\approx 1$ and a mean photon number $\langle n_\mathrm{ph}\rangle>1$ is achieved at $P\approx 0.03/$ps. In agreement with the data obtained from experiment, the coherence time reveals a slight increase at the onset of lasing. For a laser with gain provided by a QD-ensemble, coherence times of about 1\,ns are characteristic \cite{ates_influence_2008}. The much shorter coherence times observed here nicely reflect the small amount of stimulated emission provided by the combined single-emitter gain and few-emitter background gain, and the sizable impact of spontaneous emission on the above-threshold emission characteristics. We point out that the very good qualitative agreement between microscopic theory and experiment is obtained by extracting the crucial system parameters on the basis of Eq.~\eqref{eq:4D_g1_om} and consistently using these in the microscopic model. Finally, we note that the the laser threshold is crossed \emph{before} the poles merge (indicated by the vertical lines in panels (b)--(d), suggesting that SC is maintained in the presence of lasing in our device.

\section{V. Conclusion}
In conclusion, our analytical model for the strong-coupling spectrum allows for a realistic evaluation and characterization of experiments close to or at the laser threshold. In this regime, it strongly deviates from textbook equations that fail due to the onset of stimulated emission. While for a single emitter lasing in the presence of SC requires $g/\kappa$ ratios exceeding 2, for QD-microcavity systems, SC can prevail also if lasing is driven by cavity-feeding of background emitters as we have demonstrated for a QD-micropillar laser. At the same time, our results may initiate studies in systems that allow for a larger light-matter coupling, such as in superconducting-circuit QED \cite{wallraff_strong_2004}.
%

\begin{acknowledgments}
\section{acknowledgments}
The research leading to these results has received funding from the European Research Council under the European Union’s Seventh Framework ERC Grant Agreement No. 615613, and from the German Research
Foundation via projects Re2974/10-1, Gi\mbox{1121/1-1}, JA619/10-3, JA619/13-1. P.G. acknowledges financial support from the PNII-ID-PCE Research Program (Grant Nr.103/2011) and the Core Program PN16-480101. We gratefully acknowledge expert sample preparation by M.~Emmerling and thank L.~Me{\ss}ner for technical assistance.
\end{acknowledgments}

\section{Appendix}
\subsection{1. Details of the model with background emitters}

To account for exciton and higher multiexciton states commonly found in solid-state QD emitters, we typically model each QD by considering several confined single-particle states for electrons and holes \cite{gies_single_2011}. For such a system, the increasing size of the Hilbert space with emitter number limits calculations to $\approx 4$ QDs \cite{florian_phonon-mediated_2013}. To be able to evaluate the equations for more emitters, we use an effective model, where we consider only the resonant transitions with the cavity mode of each emitter, where each transition is then described in terms of a two-level system. For QD~1, of which the exciton transition is strongly coupled to the mode, the two-level system accounts for the exciton to ground-state transition that is driven by the pump rate $P_1=P$.

From the constant antibunching observed in the experiment at low excitation, we conclude that background effects appear only at elevated pumping. While detuned emitters are excited by the incoherent pumping, their exciton transition is too far detuned from the cavity mode to be coupled. As it has been shown in \cite{laucht_temporal_2010, winger_explanation_2009}, large detunings to the mode are easily bridged by the appearance of dense-lying multi-exciton states. In our model, we account out of a multitude of possible transitions for one that is resonant with the mode. We consider up to seven additional background emitters. For these QDs 2-8, the two levels then account for the transition between this multi-exciton state and a state from the manifold of multi-exciton states with one excitation less. At high excitation, these multi-exciton transitions are mainly responsible for the emission properties and properly treated in our formalism. 

To describe the switch-on behavior of the detuned emitters in the regime of intermediate excitation powers, we use a simple phenomenological model that accounts for (1) the higher-order pump dependence typical for multi-exciton transitions at least $\propto P^2$ and (2) the presence of multi-exciton states if the emitter are driven beyond the saturation of the exciton transition $P^\textrm{X}_{\textrm{sat}}$. Above, we use $P_{2-8} =\alpha P^2$ with $\alpha<1$. Below, carrier occupations in the background emitters are too low for multi-exciton states to form, and their effect is negligible. We point out that the exact way how background-emitters begin to contribute in this transition region is not our topic of investigation and neither in the low, nor in the high-excitation regime are the numerical results influenced by this procedure.



We point out that our model is based on explicit assumptions on the experimental situation, which due to limited computational resources are impossible to model on a fully microscopic level. Nevertheless, it captures the main elements of a single-QD microcavity system in the strong coupling regime and in the presence of detuned background emitters that provide additional gain required to reach lasing, and it does so under full consideration of the light-matter interaction required to model the coherent strong-coupling regime. Thereby, we can offer an interpretation of the physics underlying the experiment that is in excellent qualitative agreement with several observables at the same time.

\begin{figure}
\centering
\includegraphics[width=0.9\columnwidth]{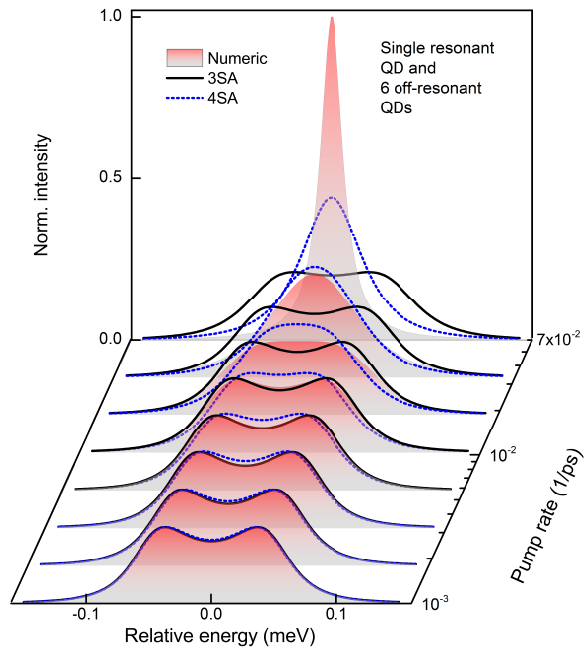}
\caption{Spontaneous emission spectra obtained from a solution of Eq.~\eqref{eq:vNL} in the main text for a resonant single QD and 6 detuned background emitters (solid lines), together with the analytic expressions used in Fig.~2(c) in the main text. Parameters as in Fig.~2(c) with an additional homogeneous dephasing of $\gamma_h=30\,\mu$eV. All shown spectra are normalized to unity area.}
\label{fig:7QDspec}
\end{figure}

The spontaneous emission spectra that are obtained from the numerical solution of the von Neumann equation for the single QD plus 6 background emitters are shown in Fig.~\ref{fig:7QDspec}. A homogeneous dephasing of 30$\,\mu$eV has been added \cite{ortner_temperature_2004} to match the situation in the experiment, which is performed at 25K. The coherence time shown in Fig.~3(e) has been obtained from these spectra. The increase in coherence time due to the onset of lasing is reflected in a linewidth narrowing at the highest pump rate, which is absent in the single-QD case shown in Fig.~2 in the main text.


\subsection{2. Comparison of spectra in the different approximations}

\begin{figure}[h!]
\includegraphics[width=\columnwidth]{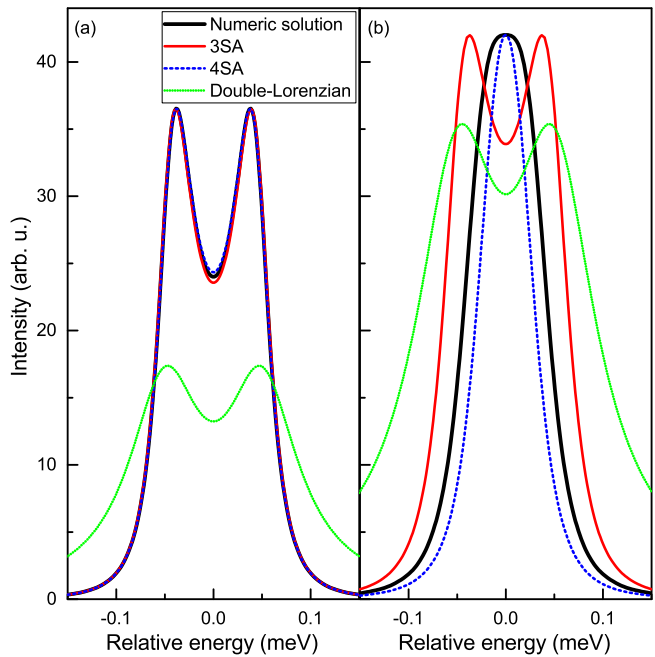}
\caption{Comparison of different fits to the numerically calculated cavity emission spectrum at \textbf{a)} low and \textbf{b)} high excitation.}
\label{fig:spec_vergleich}
\end{figure}	

Spectra in the strong coupling regime are often described in terms of a superposition of two Lorentzian lines. Both for assessing  the relevance of the strong coupling regime in a given situation, and in order to correctly fit experimental data, it is important to use a model that captures the essential physics in that regime. The approximation of two separate Lorentz peaks is only applicable under weak excitation and in an environment, where the light-matter coupling strength greatly exceeds the strength of any dissipative channels. Therefore, it is generally not suited for driven QD-microcavity systems, where excitation-induced dephasing alone can be significant.

To illustrate the difference between various models in use, we compare in Fig.~\ref{fig:spec_vergleich} the numerical spectrum for a single QD and the parameters used in Fig.~2 in the main text to different analytical expressions for the two cases of weak and strong excitation. At low excitation the full numerical solution of the von Neumann equation (Eq.~\eqref{eq:vNL}) (black) is well described by the commonly used 3SA (red). Considering an additional state in the 4SA (dashed) only leads to a minor correction. The situation is very different at high excitation, where the 3SA and 4SA differ completely in their prediction about strong coupling and peak splitting. In using the commonly used 3SA in such a regime as a fit to experimental data, one would obtain parameters that do not correctly relate to the experiment. Interestingly, this deviation between the 3SA and 4SA is not related to pump-induced dephasing, which is accounted for in both cases, but arises from the truncation of the Hilbert space.

A fit using two Lorenzian lines (green) is inaccurate even at low excitation due to the presence of dissipation in the QD-microcavity system.

\subsection{3. Fits to the experimental spectra}

\begin{figure}[]
 \centering
  \includegraphics[width=\columnwidth]{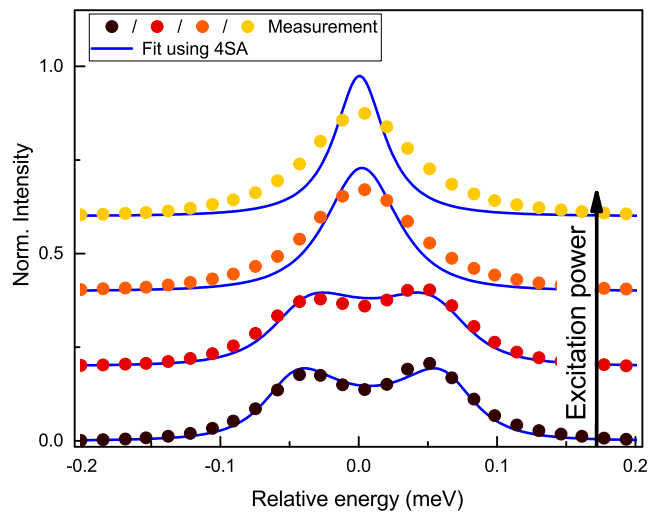}
  \caption{Selection of the measured spectra shown in Fig.~\ref{fig:thexp}(a) of the main text, together with the fits as obtained from the 4SA.}
  \label{fig:fitspectra}
\end{figure}

We used a least square optimization to fit the model of the 4SA to the data to estimate the coupling constant $g$. To convert the measured power to a pump rate we assumed a linear dependence: $P_{1/ps} = \alpha_{power} \cdot P_{W}$. The parameter $\alpha_{power}$ can only be fitted to the data when we fit all the spectra of the power dependant measurement at once. That means that the fit parameter for: $g$, $\gamma_h$, $\gamma$, $\kappa$, $\alpha_{power}$ were kept the same for all spectra. Only a scaling factor for the intensity and an offset for the central position of the peak were varied from spectra to spectra. We introduced the individual scaling factor for the intensity to take the off-resonantly coupled QDs into account, because the 4SA was derived for only a single QD in resonance with the cavity mode. To limit the number of free parameters even further we estimated $\kappa$ from the linewidth of the cavity separatly at high excitation power with no particular QD tuned in resonance. The value for $\gamma_h$ was taken from \cite{ortner_temperature_2004} as a typical value for QD emitter pure dephasing at 25K.
Exemplary fits to the measured spectra throughout the whole excitation range are shown in Fig.~\ref{fig:fitspectra}.

\subsection{4. Temperature tuning of the emission from the QD-micropillar system}
\begin{figure}[b]
 \centering
  \includegraphics[width=\columnwidth]{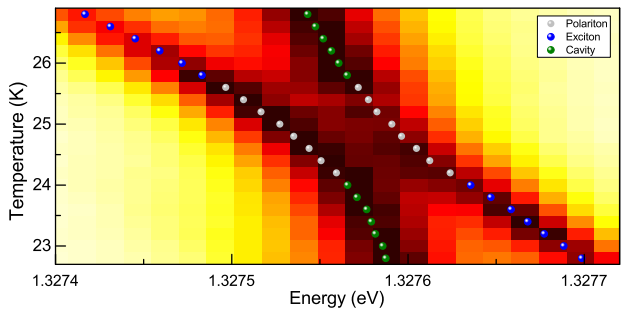}
  \caption{Temperature dependence of low-excitation $\mu$PL spectra showing the tuning of the single QD exciton through resonance of the micropillar cavity, performed at  at 0.04$\,\mu$W excitation power.}
  \label{fig:temperaturecurves}
\end{figure}

In Fig.~\ref{fig:temperaturecurves} we show results under temperature tuning of the QD-micropillar at low excitation energies. A clear anti-crossing of a single-QD exciton (X) and the fundamental cavity mode (C) with a vacuum Rabi splitting of about $60\,\mu$eV is revealed at the resonance temperature of 24.6\,K. Experimental results shown in the main text have been obtained at the resonance temperature. To compensate for laser-induced heating and to maintain the resonance condition, the temperature of the sample was slightly readjusted during the measurements.

\subsection{5. Autocorrelation measurements}

HBT autocorrelation measurements are limited by the total temporal resolution of the setup, which must be sufficient to resolve the autocorrelation function with respect to the delay time $\tau$ between two emission events. In the regime of antibunching, the $\tau$ dynamics takes place on the timescale of the emitter's cavity-enhanced emission lifetime, whereas in the thermal regime, it is determined by the coherence time as shown in Fig.~3(b) and (e) in the main text.

The measured signal ($g^{(2)}_{meas}$) consist of the ideal signal ($g^{(2)}_{ideal}$) that is convoluted with a Gaussian function with the area normalized to one and a width (full width at half maximum) of the total temporal resolution. To determine the $g^{(2)}(\tau = 0)$, we have to fix the lifetime or coherence time. Hence we estimate the range of the expected lifetimes and coherence times. The spontaneous lifetime ($\tau_l$) of a QD exciton into a cavity mode can be calculated by \cite{gies_semiconductor_2007}
$$
\tau_l = \frac{ \kappa + \Gamma }{ 2 g^2 }.
$$
For the parameters discussed above, we estimate a lifetime of $\tau_l \approx 10\,$ps. Measurements performed on cavity structures featuring comparable values for the Q-factor and $g$, show a lifetime of about $\tau_l \approx 20\,$ps \cite{muller_ultrafast_2015} for a QD exciton in the strong coupling regime. In the case of a QD spectrally detuned from the cavity mode, the lifetime increases with respect to the resonant case. The maximum coherence time was measured to be on the order of $35\,$ps (Fig.~3(b) in the main text). We expect the lifetime and coherence time of our strongly coupled QD to be in the range of $10\,$ps to $35\,$ps.


\bibliography{scl_2016,scl_2016_addon}

\end{document}